\documentclass[twocolumn,aps,prl,superscriptaddress]{revtex4-1}
\usepackage{graphicx}
\usepackage{soul}
\usepackage{xcolor}

\begin{document}

\title{Ferromagnetic Kitaev interaction and the origin of large magnetic anisotropy in $\alpha$-RuCl$_3$}

\author{J. A. Sears}
\affiliation{Deutsches Elektronen-Synchrotron (DESY), 22607 Hamburg, Germany}

\author{Li Ern Chern}
\affiliation{Department of Physics, University of Toronto, 60 St.~George St., Toronto, Ontario, M5S 1A7, Canada}

\author{Subin Kim}
\affiliation{Department of Physics, University of Toronto, 60 St.~George St., Toronto, Ontario, M5S 1A7, Canada}

\author{P. J. Bereciartua }
\affiliation{Deutsches Elektronen-Synchrotron (DESY), 22607 Hamburg, Germany}

\author{S. Francoual}
\affiliation{Deutsches Elektronen-Synchrotron (DESY), 22607 Hamburg, Germany}

\author{Yong Baek Kim}
\affiliation{Department of Physics, University of Toronto, 60 St.~George St., Toronto, Ontario, M5S 1A7, Canada}

\author{Young-June Kim\thanks{yjkim@physics.utoronto.ca}}
\affiliation{Department of Physics, University of Toronto, 60 St.~George St., Toronto, Ontario, M5S 1A7, Canada}

\date{\today}

\maketitle

\textbf{$\alpha$-RuCl$_3$ is drawing much attention as a promising candidate Kitaev quantum spin liquid \cite{plumb2014,sears2015,sandilands2015, johnson2015, banerjee2016, banerjee2017, do2017,kasahara2018}. However, despite intensive research efforts, controversy remains about the form of the basic interactions governing the physics of this material. Even the sign of the Kitaev interaction (the bond-dependent anisotropic interaction responsible for Kitaev physics) is still under debate, with conflicting results from theoretical and experimental studies \cite{hskim15,banerjee2016,banerjee2017,hskim16,winter2016,yadav2016,hou2017,wang2017,eichstaedt2019}. The significance of the symmetric off-diagonal exchange interaction (referred to as the $\Gamma$ term) is another contentious question \cite{rau2014,katukuri2014,chaloupka2015}. Here, we present resonant elastic x-ray scattering data that provides unambiguous experimental constraints to the two leading terms in the magnetic interaction Hamiltonian. We show that the Kitaev interaction ($K$) is ferromagnetic, and that the $\Gamma$ term is antiferromagnetic and comparable in size to the Kitaev interaction. Our findings also provide a natural explanation for the large anisotropy of the magnetic susceptibility in $\alpha$-RuCl$_3$ as arising from the large $\Gamma$ term. We therefore provide a crucial foundation for understanding the interactions  underpinning the exotic magnetic behaviours observed in $\alpha$-RuCl$_3$.}

The magnetic behaviour of the honeycomb material $\alpha$-RuCl$_3$ has been the topic of much recent work, following the discovery in this material of an unusual continuum of magnetic excitations not well explained by spin-wave theory \cite{sandilands2015, banerjee2016, banerjee2017}. The structural environment and electronic state of the ruthenium atoms in $\alpha$-RuCl$_3$ are such that the Kitaev magnetic interaction \cite{kitaev2006} is expected to be significant \cite{jackeli2009,plumb2014}. For this reason, these remarkable findings have been attributed to fractionalized excitations analogous to those found in the spin liquid ground state of the Kitaev model \cite{banerjee2016, banerjee2017}. Recent discovery of quantization of the thermal Hall signal in the intermediate magnetic field phase has further stimulated interest in this material \cite{kasahara2018}.

Understanding the salient features of magnetism in $\alpha$-RuCl$_3$ requires a good knowledge of the magnetic interactions between the ruthenium magnetic moments. The magnetic Hamiltonian relevant to this material includes an isotropic Heisenberg ($J$) term as well as bond-dependent anisotropic Kitaev ($K$), and off-diagonal $\Gamma$ terms \cite{rau2014}. The general Hamiltonian for atoms at adjacent sites $i$ and $j$ takes the following form, where $\alpha$, $\beta$, and $\gamma$ denote the spin components:
\begin{equation}
\mathcal{H}_{ij}^{(\gamma)} = J \mathbf{S}_i \cdot \mathbf{S}_j + K S^{\gamma}_i S^{\gamma}_j + \Gamma (S^{\alpha}_i S^{\beta}_j + S^{\beta}_i S^{\alpha}_j).
\end{equation}
Often included in this Hamiltonian are further-neighbour isotropic interactions ($J_2$, $J_3$, etc.) and additional off-diagonal term $\Gamma^\prime$ due to non-zero trigonal crystal fields \cite{rau2014,katukuri2014,yamaji2014}. Early \textit{ab initio} calculations \cite{hskim15} and fits to inelastic neutron scattering measurements \cite{banerjee2016,banerjee2017} suggested an antiferromagnetic Kitaev interaction ($K>0$). However, later calculations using the updated monoclinic crystal structure \cite{johnson2015} instead suggested that the Kitaev term is ferromagnetic ($K<0$) \cite{hskim16,winter2016,yadav2016,hou2017,wang2017,eichstaedt2019}.

Although these scenarios can be distinguished by the direction of the ordered magnetic moment \cite{chaloupka2016}, to date, this information has not been experimentally available. The magnetic structural solution from neutron diffraction data \cite{cao2016} suggested two possible structures, with collinear moments confined to the monoclinic ac plane (See Fig. 1a ). These two magnetic structures, differing only in the canting angle of the moment direction out of the crystallographic ab (honeycomb) plane ($\Theta$), were fit equally well by the neutron data. In the case of $K>0$, the moment is expected to point along the local cubic axis (towards a Cl atom), and therefore $\Theta=-35^{\circ}$, while $\Theta=+35^{\circ}$ corresponds to a ferromagnetic $K$ ($K<0$).

\begin{figure*}
\includegraphics[width=6.5in]{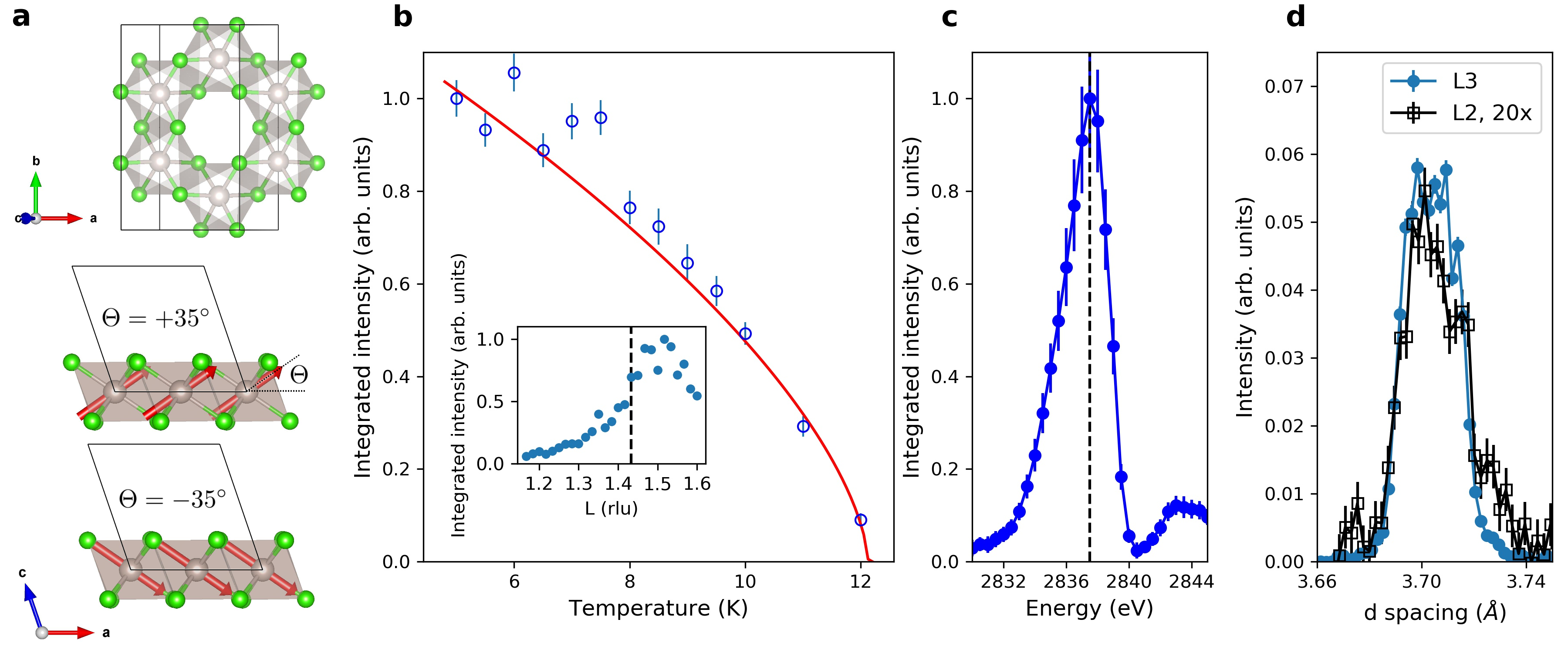}
\caption{{\bf Characterization of magnetic scattering. a.} Crystal structure and ordered moment directions of $\alpha$-RuCl$_3$ proposed in Ref.~\cite{cao2016}. {\bf b.} Temperature dependence of magnetic diffraction intensity at (0,-1,1.43), showing an ordering temperature of 12K. At each temperature the magnetic peak was measured by simultaneously scanning the sample and detector angles. Inset: intensity dependence on L reciprocal space direction (perpendicular to the honeycomb plane) showing a broad peak at the L=1.5 position. At each L value, the magnetic peak was measured by scanning along the momentum space K direction. The integrated intensities for all scans were found by fitting the scans with a Gaussian peak shape. Error bars shown are the square root of covariance value from the fit. {\bf c.} Energy dependence of the magnetic diffraction intensity at (0,-1,1.43), showing resonance at the Ru $L_3$ resonant energy of 2837.5 eV. Integrated intensities and error bars were calculated from combined scans of the sample and detector angles, as in {\bf b.} {\bf d.} Comparison of the magnetic signals obtained with the incident photon energy at the Ru $L_3$ edge (2837.5 eV) and the Ru $L_2$ edge (2970 eV). Scans were collected by simultaneously scanning sample and detector angles. Error bars shown are the square root of the number of photons detected. A constant background was subtracted, and the overall photon counts normalized to the monitor recording incident beam intensity.}
\label{fig1}
\end{figure*}

This ambiguity can be resolved with resonant elastic x-ray scattering (REXS). The magnetic scattering process in x-ray diffraction is fundamentally different from that of neutron diffraction, allowing the moment direction to be determined by measuring the {\em azimuthal} dependence of a magnetic Bragg peak intensity. This measurement is done by rotating the sample around the scattering vector ($\vec{q}$) as shown in Fig. 2a. With the incident linear polarization perpendicular to the vertical scattering plane, the diffracted magnetic intensity for electric dipole transitions is proportional to the projection of the ordered moment onto the scattered beam \cite{hill1996} and shows modulation as the sample is rotated about the scattering vector. By modelling this intensity modulation as a function of the azimuthal angle, $\Psi$, one can distinguish between the two possible structures suggested by the neutron measurement.

We have collected REXS data on a single crystal sample of $\alpha$-RuCl$_3$ at the known magnetic Bragg peak position expected for zigzag magnetic ordering \cite{banerjee2016, sears2015}. The magnetic diffraction signal in this sample was first characterized by measuring its dependence on momentum, temperature, and incident photon energy. The momentum dependence of the magnetic signal showed an extended rod in the out-of-plane direction, with a broad peak at the position expected for ABAB type layer stacking as shown in Fig. 1b (inset). This stacking order has previously been reported in neutron diffraction measurements with an ordering temperature of 14 K, as opposed to the 7 K ordering temperature observed for the three-layer stacking \cite{banerjee2016}. We measured an ordering temperature of 12 K for this sample, as shown in Fig. 1b. We note that diffraction measurements at this x-ray energy will be highly surface-sensitive, since the beam penetrates the sample to a depth of only a few hundred nm, and the observed 2-layer stacking may not reflect the bulk crystal structure. As the magnetic interactions are strongly two-dimensional, we do not anticipate that stacking has a large effect on the moment direction. Scans along the L direction in reciprocal space were also collected at several different azimuthal positions, to ensure that the azimuthal dependence does not depend on the L position selected. The position L=1.43 was selected to maximize intensity while maintaining an accessible position for the diffractometer.

The azimuthal dependent measurement was collected at an incident photon energy of 2837.5 eV (corresponding to the Ru $L_3$ edge), where the intensity is at a maximum. The dependence of the peak intensity on the incident photon energy was measured both to find the optimal energy for measurement, and to confirm the resonant nature of the magnetic peak. This energy dependence is plotted in Fig. 1c. Following the measurements at the $L_3$ edge, the magnetic peak intensity at the same reciprocal space position was also measured at the ruthenium $L_2$ edge (incident photon energy 2970 eV). The integrated intensity was substantially weaker at the $L_2$ edge (comparison is shown in Fig. 1d), and we calculate a ratio of approximately 20 for the intensities at the two photon energies.

\begin{figure*}
\includegraphics[width=6.5in]{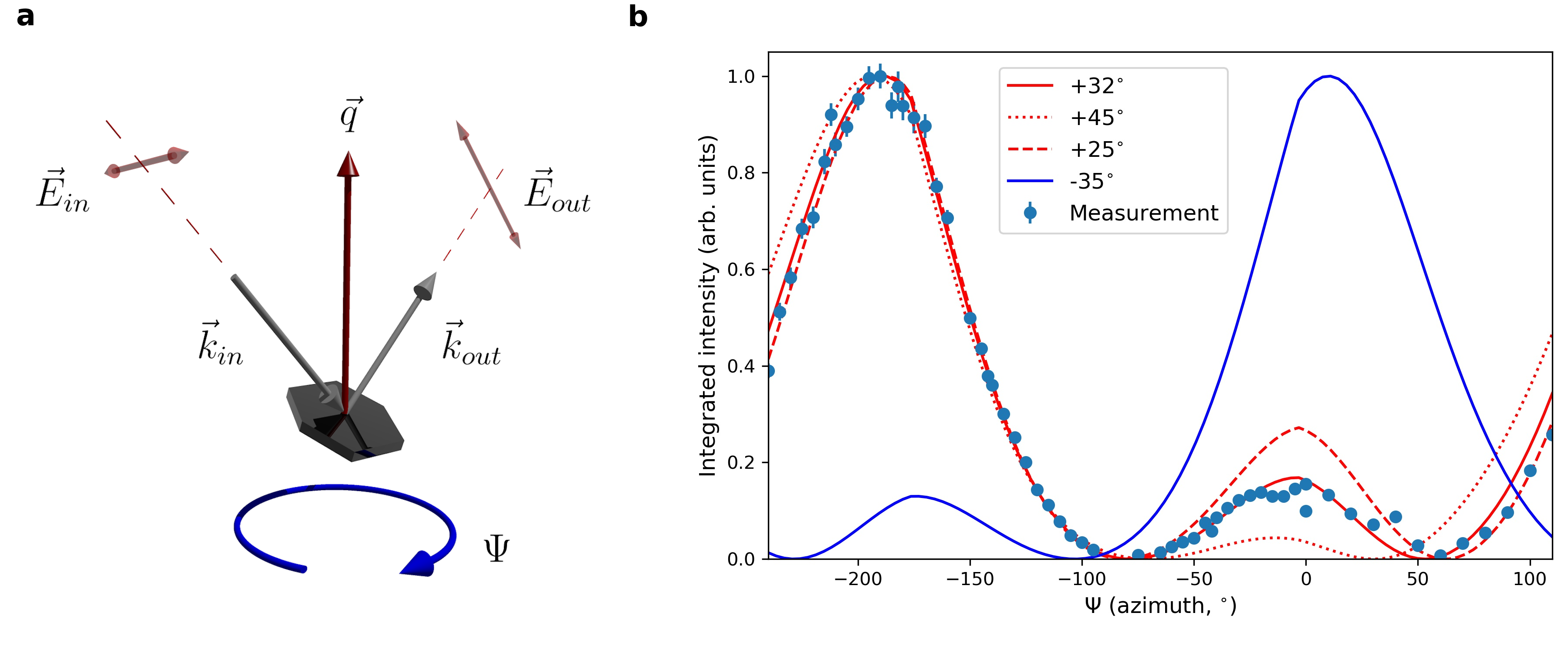}
\caption{{\bf Azimuthal dependence. a.} Schematic diagram showing the geometry of the REXS experiment. {\bf b.} Azimuthal dependence of the magnetic diffraction signal at (0,-1,1.43). The azimuthal dependence is fit best with a magnetic moment angle of $\theta=+32^{\circ}$. The modeled intensities for $\Theta=+25^{\circ}$, $+45^{\circ}$ and $-35^{\circ}$ are shown for comparison. $\Psi=0$ corresponds to the position with the in-plane direction (-2,0,0.65) pointing along the scattered beam. The magnetic peak was measured by scanning the sample angle. Integrated intensities were found by fitting the scans with a Gaussian peak shape. Error bars shown are the square root of the covariance value from the fit.}
\label{fig2}
\end{figure*}

The detailed azimuthal dependence of the magnetic scattering measured at the (0,-1,1.43) reciprocal lattice position is shown in Fig. 2b, which exhibits substantial variation in intensity as the sample is rotated about $\vec{q}$. The zero position of the azimuthal angle $\Psi$ corresponds to the orientation with the in-plane direction (-2,0,0.65) pointing along the outgoing beam. The maxima in intensity correspond to the positions when the moment lies in the scattering plane, while the minima are at positions where the magnetic moment lies approximately orthogonal to the scattering plane. The difference in intensity of the two maxima is directly related to the degree of out-of-plane canting, with the largest peak directly indicating which way the moment is canted out of the honeycomb plane. This can be seen in the modeled intensity for the two proposed moment directions (Fig. 2b), which show opposite behaviour in this respect. The measured azimuthal dependence collected for $\alpha$-RuCl$_3$ is clearly fit best by the model with the moment direction pointing towards the RuCl$_6$ octahedral face, indicating that the moment direction is along the face-centered direction expected in the case of a ferromagnetic Kitaev term.

We also allow the angle within the ac plane ($\Theta$) to vary as shown in dashed lines in Fig. 2b. The best fit is obtained when the moment is confined to the ac plane, with $\Theta = 32^\circ \pm 3^\circ$. This result is consistent with one of the two models proposed by the neutron diffraction result, and also provides insight into the form of the magnetic Hamiltonian. In the case of a ferromagnetic K term, Chaloupka and Khaliullin showed that a substantial antiferromagnetic $\Gamma$ interaction term is required to keep the moment in the ac plane. Specifically they showed that with increasing $\Gamma$, the moment rotates away from the local octahedral xy plane ($\Theta \sim 50^{\circ}$) and slowly approaches $\Theta = 32^{\circ}$ from the positive side. According to Ref.~\cite{chaloupka2016}, in order to have $\Theta \sim 32^{\circ}$ the magnitude of $\Gamma$ must be a significant fraction of, or even exceed the magnitude of $K$. We note that $\Theta \sim 45^{\circ}$ was obtained for another Kitaev material Na$_2$IrO$_3$ \cite{chun2015}, which would suggest that the $\Gamma$ term is much smaller in Na$_2$IrO$_3$.

Our REXS results provide a clue for solving one of the remaining questions regarding the magnetic properties of $\alpha$-RuCl$_3$: its large magnetic anisotropy. As reported by many groups \cite{sears2015,kubota2015,majumder2015,johnson2015}, the in-plane magnetic susceptibility measured by applying magnetic field along the direction in the ab plane is significantly larger than the out-of-plane susceptibility. A conventional way to explain this would be resorting to the $g$-factor anisotropy. However, experimental data suggest that $g$-factor anisotropy cannot be very large, certainly not large enough to account for the anisotropic susceptibility \cite{agrestini2017,lampenkelley2018}. Another route to obtain a large magnetic anisotropy is via a large $\Gamma$ term as suggested in Ref.~\cite{gohlke2018}. Physically, the effect of the $\Gamma$ interaction is to force the moments towards the ab plane, which accentuates magnetic anisotropy.

\begin{figure}
\includegraphics[width=3.5in]{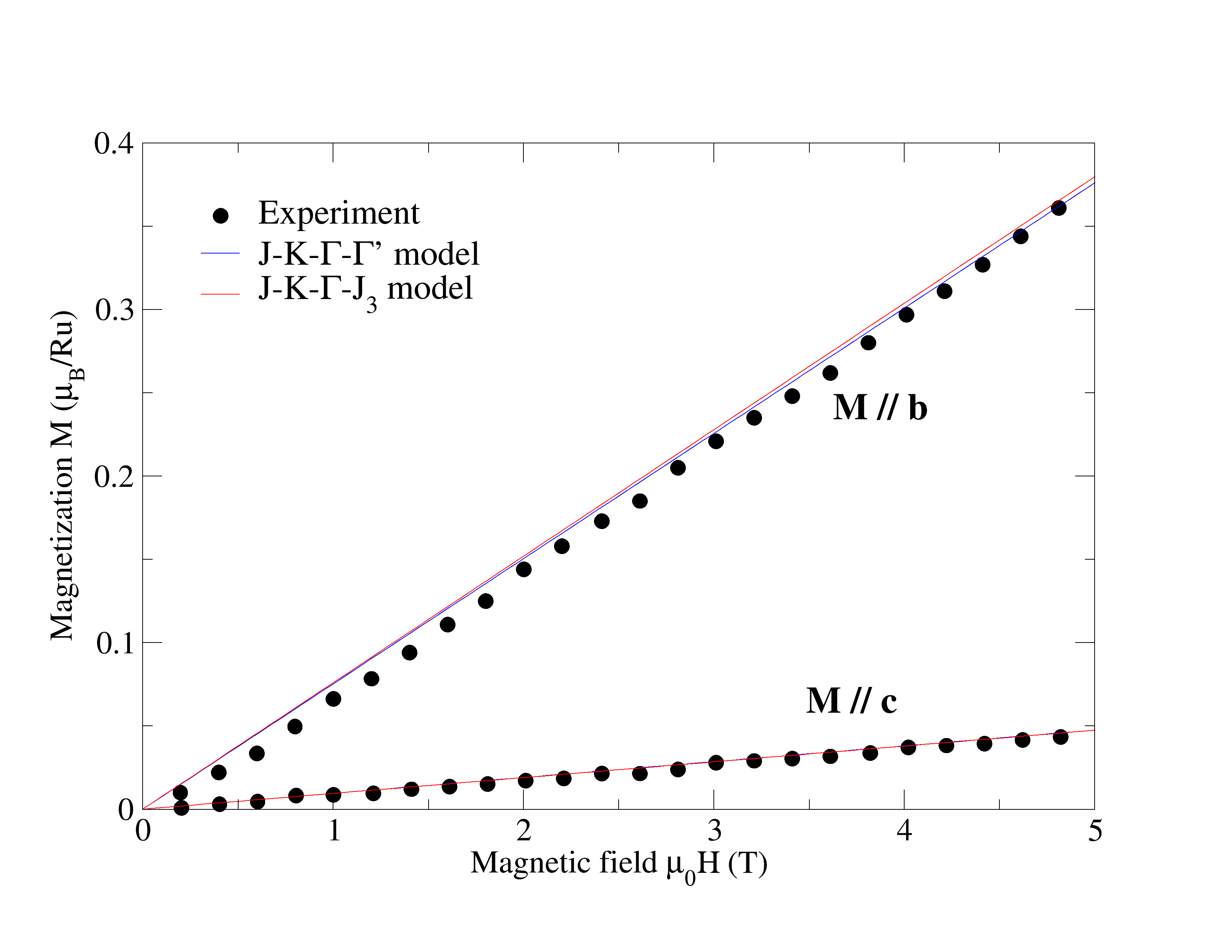}
\caption{Fitting the experimental data through simulated annealing calculations on the classical spin model. The experimental measurements were carried out using a commercial SQUID magnetometer at 2 K. The two representative parametrizations theory-I and theory-II correspond to $(J,K,\Gamma,\Gamma')=(-2.7,-10,10.6,-0.9)$ and $(J,K,\Gamma,J_3)=(-1.5,-10,8.8,0.4)$, which yield the angles $\Theta=33.3^{\circ}$ \, and $\Theta=34.3^{\circ}$ \, between the moments and the honeycomb plane at zero field, respectively. Energy is in units of $\mathrm{meV}$. The magnetization curves of these two parameterizations are very similar such that they overlap each other. We fix $g=2.3$ and $S=1/2$ throughout the calculations.}
\label{fig3}
\end{figure}

We demonstrate that a large $\Gamma$ is sufficient to explain the observed magnetic anisotropy by comparing the experimental data with theoretical calculation results. The low-field magnetization data for fields applied in-plane and out-of-plane are plotted in Fig.~3, which shows that the susceptibility (slope) anisotropy is about $\chi_{ab}/\chi_c \sim 8$. This data is fit with the classical $J K \Gamma$ model (Eq.~(1)), where the model parameters are chosen to be consistent with the magnetic moment direction determined by REXS. Either a small $\Gamma^\prime$ or $J_3$ term was added to ensure the zigzag ground state of the model (details about the calculation are provided in the Supplementary Information). The data can be fitted for several parameter choices with ferromagnetic $K$ and antiferromagnetic $\Gamma$ of similar magnitude, demonstrating that the magnetization data can be explained without resorting to $g$-factor anisotropy.  We note that in \cite{gohlke2018} it was shown that a ratio of $|\Gamma/K| \sim 1$ can also explain the star-shaped continuum intensity centered around the Brillouin zone center observed in inelastic neutron scattering \cite{banerjee2017}.

The measurements outlined in this paper have determined that the ordered moment direction in $\alpha$-RuCl$_3$ points toward the octahedral face, rather than towards one of the cubic axes of the RuCl$_6$ octahedra. This result establishes that the Kitaev interaction is ferromagnetic in this material. In addition, we show that a substantial antiferromagnetic $\Gamma$ interaction is essential for understanding magnetism of $\alpha$-RuCl$_3$. In particular, the presence of large $\Gamma$ interaction could reconcile the large magnetic anisotropy observed experimentally with the almost isotropic $g$-factors expected in this material. The findings of our REXS measurement provide new experimental constraints on the magnetic Hamiltonian of $\alpha$-RuCl$_3$, indicating that it lies within the ferromagnetic K, antiferromagnetic $\Gamma$ regime \cite{chaloupka2016}. This result is in agreement with the findings of a number of ab-initio calculations, and can inform future investigations into the unusual magnetic behavior of $\alpha$-RuCl$_3$.

\section*{Methods}
\setlength{\parskip}{12pt}

REXS measurements were carried out at the beamline P09 at PETRA III at DESY (see \cite{strempfer2013} for details) at the ruthenium $L_3$ and $L_2$ edges (2838 and 2967 keV respectively). Most of the measurements, including momentum, temperature, and azimuthal dependence were collected at the $L_3$ edge. The magnetic intensity was also measured at the $L_2$ edge to determine the branching ratio. The monochromator was detuned to minimize the presence of higher harmonics in the beam, and the measurements were made with a sodium iodide scintillation detector. An all-in-vacuum path was used to minimize x-ray absorption by air. $\alpha$-RuCl$_3$ single crystals were grown by vacuum sublimation in sealed quartz tubes using commercial RuCl$_3$ powder. The single crystal used for this measurement was a flat plate with largest dimension $\sim$400 $\mu$m.

The orientation of the crystal used for this measurement was checked at room temperature with the crystal in the known monoclinic structural phase, by checking structural Bragg peaks using higher energy (third harmonic) photons. The crystal was also checked to ensure that it did not possess a twin rotated by 180$^{\circ}$, which would affect the result of the azimuthal measurement. This was done by searching for structural peaks at the positions expected for the rotated structure. No intensity was found at the peak positions expected for the rotated crystal structure. The azimuthal dependence data was corrected for beam absorption (as described in \cite{bruckel2001}), and the beam footprint on the sample. Beam footprint on the sample was calculated from the angle between the sample surface and the incoming beam, and depended only on the ratio of the beam height and the smallest dimension of the sample. This ratio of beam height to sample size, and the magnetic moment angle $\Theta$ were the only parameters refined in the fit of the azimuthal dependence data. More detailed information about the fitting procedure can be found in the Supplementary Information.

\section*{Acknowledgements}
\setlength{\parskip}{12pt}

We would like to thank Joel Bertinshaw and Hakuto Suzuki for their help with the experiment. We
acknowledge DESY (Hamburg, Germany), a member of the Helmholtz Association HGF, for the provision of experimental facilities. Parts of this research were carried out at PETRA III.
Work at the University of Toronto was supported
by the Natural Science and Engineering Research
Council (NSERC) of Canada, Canadian Foundation
for Innovation, Ontario Innovation Trust, and the Center for Quantum Materials at the University of Toronto.
Y.B.K. is also supported by the Killam Research Fellowship
from the Canada Council for the Arts.
This work was performed in part at Aspen Center for
Physics, which is supported by National Science
Foundation grant PHY-1607611.



\end{document}